\documentstyle[preprint,floats,pra,aps]{revtex} 
\tightenlines  
\begin{document} \draft 

\def\be{\begin{equation}}
\def\ee{\end{equation}}
\def\bea{\begin{eqnarray}}
\def\eea{\end{eqnarray}}

\title{\LARGE \bf B\"{a}cklund-type superposition and free
particle $n$-susy partners}

\author{O. Rosas-Ortiz${}^{1,2}$, B. Mielnik${}^{1,3}$, and L.M.
Nieto${}^2$\\[1ex]}

\address{${}^1$Departamento de F\'{\i}sica, CINVESTAV-IPN, AP 14-740,
M\'exico 07000, DF, Mexico\\
${}^2$Departamento de F\'{\i}sica Te\'orica, Universidad de Valladolid,
47011 Valladolid, Spain\\
${}^3$Institute of Theoretical Physics, Warsaw University, H\.{o}za 69
Warsaw, Poland}

\maketitle 
\thispagestyle{empty}

\begin{abstract}
The higher order susy partners of Schr\"odinger Hamiltonians can be
expli-citly constructed by iterating a nonlinear difference algorithm
coinciding with the B\"{a}cklund superposition principle used in soliton
theory.  As an example, it is applied in the construction of new higher
order susy partners of the free particle potential, which can be used as a
handy tool in soliton theory.
\end{abstract} 

\vspace{8mm} 


Recent studies confirm that the higher order supersymmetric (susy) 
partners of Schr\"{o}dinger Hamiltonians are most easily constructed by a
simple algebraic tool named intertwining technique [1]. One of the keys of
this method is an algebraic nonlinear expression which links solutions of
different Riccati equations (see, e.g. [2--4]).  In a previous paper [3],
we have studied the application of this method to the free particle
potential. The `building blocks' of some of the resulting potentials are
the well known soliton solutions of the Korteweg-de Vries (KdV) equation:
$\kappa^2 {\rm sech}^2 [ \kappa (x - a)]$ and $\kappa^2 {\rm csch}^2 [
\kappa (x - a)]$.  In this work we shall sketch the main steps of the
approach in order to present some of the potentials
derived in [3]. 

First, consider the intertwining relationship $H_1 A_1 = A_1 H_0$, where
the intertwiner $A_1$ is the first order differential operator $A_1=
\frac{d}{dx} + \beta_1(x, \epsilon)$. All the available information
concerning the Hamiltonians $H_0 = -\frac{1}{2} \frac{d^2}{dx^2} + V_0(x)$
and $H_1 = -\frac{1}{2} \frac{d^2}{dx^2} + V_1(x,\epsilon)$ is encoded in
the beta function, which satisfies the Riccati equation
\be 
-\beta'_1(x,\epsilon) + \beta_1^2(x,\epsilon) = 2[V_0(x) - \epsilon].
\label{riccati1}
\ee
The arbitrary integration constant $\epsilon$ plays the role of a
{\it factorization energy\/}. It is very simple to check that the
potentials are related by the first order susy relationship
\be  
V_1(x,\epsilon) = V_0(x) + \beta'_1(x,\epsilon).
\label{susy1}
\ee
Equations (\ref{riccati1}) and (\ref{susy1}) are necessary and sufficient
conditions for the Hamiltonians to be factorized as $H_0 - \epsilon =
(1/2)\, A_1^{\dag} A_1$, and $H_1 - \epsilon = (1/2)\, A_1 A_1^{\dag}$.
Suppose now that $V_0(x)$ is a known solvable potential with eigenvalues
$E_n$ and eigenfunctions $\psi_n$, $n=0,1,2,...$ Let us assume that we
have found a general solution of (\ref{riccati1}) for a given
factorization energy $\epsilon_1 \neq E_n, \forall n$. Then, the potential
$V_1(x,\epsilon_1)$ is also given [1--3]. The iteration of this
procedure starts by considering now $V_1(x,\epsilon_1)$ as the known
solvable potential and looking for a new one $V_2(x,\epsilon_1,\epsilon)$
satisfying the second order susy relationship
\be
\label{susy2}
V_2(x,\epsilon_1, \epsilon) = V_1(x,\epsilon_1) + \beta_2'(x,
\epsilon_1, \epsilon). 
\ee
Therefore, the new beta function must fulfill the Riccati equation
\be
\label{riccati2}
-\beta_2'(x, \epsilon_1, \epsilon) + \beta_2^2(x, \epsilon_1, \epsilon) =
2[V_1(x,\epsilon_1) - \epsilon],
\ee
where $\epsilon$ is again an arbitrary factorization energy. The
corresponding solution is given by
\be
\label{solucion2}
\beta_2(x, \epsilon_1, \epsilon) = -\beta_1(x,\epsilon_1) -
\frac{2(\epsilon_1 - \epsilon)}{\beta_1(x,\epsilon_1) -
\beta_1(x,\epsilon)}.
\ee 
The finite difference expression (\ref{solucion2}) is a nonlinear
superposition of two general solutions of (\ref{riccati1}), one for each
factorization energy $\epsilon_1$ and $\epsilon$, transforming equation
(\ref{riccati1}) into (\ref{riccati2}) by the change of $V_0(x)$ by
$V_1(x,\epsilon_1)$, and $\beta_1 (x, \epsilon)$ by $\beta_2
(x,\epsilon_1,\epsilon)$. This transformation can be used to link the
higher order susy partners of $V_0(x)$ with the first order
superpotentials $\beta_1 (x,\epsilon)$, just by solving (\ref{riccati1}) 
for different values of the factorization energy $\epsilon$. For instance,
providing $n$ different general solutions of (\ref{riccati1}), one for
each $\epsilon_k$, $k=1,2,...,n$, we are able to iterate $n-1$ times the
algorithm (\ref{solucion2})  acquiring a new beta function in each step,
given by
\be
\beta_k (x,\epsilon_k) = - \beta_{k-1} (x,\epsilon_{k-1}) - \frac{2(
\epsilon_{k-1} - \epsilon_k)}{\beta_{k-1} (x,\epsilon_{k-1}) - \beta_{k-1}
(x,\epsilon_k)}, \qquad k=2,3,...n.
\label{solucionk}
\ee
We have adopted here an abreviated notation making explicit only the
dependence of $\beta_k$ on the factorization constant introduced in the
very last step, keeping implicit the dependence on the previous
factorization constants (henceforth, the same criterion will be used for
any other symbol depending on $k$ factorization energies). Therefore,
given any initial potential $V_0$, the corresponding $n$-susy partner
potential $V_n$ can be writen as
\be
\label{final} 
V_n(x,\epsilon_n) = V_0 (x)  + \sum_{k=1}^{n} \beta_k'(x,
\epsilon_k),
\ee
provided that the master equations for $\beta_k$ and $V_{k}$ are given by
\be
\label{master1}
-\beta_k'(x, \epsilon_k) + \beta_k^2(x, \epsilon_k) =
2[V_{k-1}(x,\epsilon_{k-1}) - \epsilon_k], \qquad k=1,2,...,n,
\ee

\be
\label{master2}
V_{k}(x,\epsilon_{k}) = V_{k-1}(x,\epsilon_{k-1}) + \beta_{k}' (x,
\epsilon_{k}),  \qquad k=1,2,...,n. 
\ee
Now, let us stress that every general solution of the Riccati equation
(\ref{riccati1}), for a given $\epsilon$, depends on an additional
implicit integration parameter $\alpha$, hence, the process acumulates as
many of these integration parameters as many general solutions of
(\ref{riccati1}) have been used.

Observe the coincidence of our nonlinear algorithm (\ref{solucionk})  and
the Wahlquist and Estabrook superposition principle expression (see
equation (16) of [5]), derived from the B\"{a}cklund transformation (BT)
of the KdV equation $w_t= 6 w^2_x - w_{xxx}$; subscripts $t$ and $x$
denote partial derivatives. The method has been typically used to generate
new, multisoliton solutions $w_{12},..., w_{(n)}$ of the KdV equation from
a given one-soliton solution $w \equiv w_{1}$ of the same equation. It is
thus quite interesting that the validity of the same algorithm in the
intertwining problem (supersymmetry) is much easier to demonstrate without
worrying at all about the nonlinear equations! Moreover, its physical
applicability in susy seems much wider.  Thus, e.g., the singular
solutions of KdV (singular water waves) would be of marginal physical
interest. The singular potentials in the Schr\"{o}dinger equation are not!
Therefore, the possibility of reducing the $n$-th intertwining iteration
to the multiple applications of the B\"{a}cklund superposition principle
means that $n$-susy could be a universal method generating the
``multisoliton deformations" of any initial potential.

We shall now focus on the vacuum case, presenting some simplifications
which the method offers in deriving the $n$-susy partners for the
potential $V_0(x)=0$. In this case, the Riccati equation (\ref{riccati1})
has the general solution
\be
\beta_1(x,\epsilon)= -\sqrt{2\epsilon}\, \cot[\sqrt{2\epsilon}\,
(x-\alpha)],
\label{free1}
\ee
where $\alpha$ is an integration constant (in general complex). It is well
known that the superpotential (\ref{free1}) gives four different first
order susy partners of $V_0(x)=0$ by taking different values of $\epsilon$
and $\alpha$. This information is sumarized in Table~I. 

\vskip0.5cm
\begin{center}
{\footnotesize
\begin{minipage}{10.5cm}
TABLE I. The four different real superpotentials $\beta_1$ comming out
from (\ref{free1}), depending on the values of $\epsilon$ and the
integration parameter $\alpha$. In each case S means singular, R regular,
${\rm P}$ periodic, and N null. The parameters $a$ and
$b$ are arbitrary real numbers.
\end{minipage}
\medskip

\begin{tabular}{ccccc}
\cline{1-5} \\ [-2ex]
\cline{1-5}
& & & &   \\ [-1ex]
\quad Case \qquad & $\epsilon$ & $\sqrt{2\epsilon}$ &
\quad $\alpha $  & $\beta_1(x,\epsilon)$  \\ [1ex]
\hline
& & & &   \\ [-1ex]
S & \quad$\epsilon<0$ \quad &\quad $i\sqrt{2|\epsilon|}=i\kappa$
\quad& $a$ & \quad$-\kappa\, \coth[\kappa(x-a)]$  \quad \\ [1ex]
& & & &  \\ [-1ex]
R & $\epsilon<0$ & $i\sqrt{2|\epsilon|}=i\kappa$ & \quad
$\displaystyle -b-\frac{i\pi}{2\kappa}$ & $-\kappa\,
\tanh[\kappa(x+b)]$\quad \\ [1ex]
& & & &  \\ [-1ex]
${\rm P}$ & $\epsilon>0$ & $\sqrt{2\epsilon}= k$ & $a$ &
$-k\, \cot[k(x-a)]$      \\   [1ex]
& & & &  \\ [-1ex]
N & $0$ & $0$ & $a$  &  $\displaystyle -\frac1{x-a}$  \\ [2ex]
\cline{1-5}  \\ [-2ex]
\cline{1-5}
\end{tabular}
}
\end{center}
\vskip0.4cm
\noindent
As an example, notice that the regular case (R) leads to the well known
modified P\"oschl-Teller type susy partner $V_1^{R}(x,\epsilon)=
-\kappa^2{\rm sech}^2[\kappa(x+b)]$, while the null case (N) leads to the
potential barrier $V_1^{N}(x, 0)  = (x-a)^{-2}$. Now, in order to give an
example of second order susy partner potentials $V_2(x,\epsilon)$, let us
consider the superpotentials R and S as given in Table~I. By introducing
them in (\ref{solucion2}) and (\ref{susy2}) we get
\be
\label{free2}
V_2 (x,\epsilon_2) = - (\kappa_1^2 - \kappa_2^2) \, \, \frac{
\kappa_1^2 \, \, {\rm csch}^2 \, [ \kappa_1 (x +b)] + \kappa_2^2 \, \,
{\rm sech}^2 \, [ \kappa_2 (x -a) ] }{ (\, -\kappa_1 \coth \, [ \kappa_1
(x +b)  ] + \kappa_2 \tanh \, [ \kappa_2 (x -a) ] \,)^2 }.
\ee
The potential (\ref{free2}) has two finite wells which can be modulated 
by changing the values of $\kappa_1$ and $\kappa_2$ under the condition
$\kappa_2 < \kappa_1$. A Taylor expansion of (\ref{free2}) shows a
singularity at $x=a$ when $\kappa_2 > \kappa_1$. The case $\kappa_2 =
\kappa_1$ gives a potential $V_2 (x,\epsilon_1)=0$.

Let us remark that, for the periodic superpotentials $\beta_1$ in Table~I,
equation (\ref{final}) leads to a natural classification of two kinds of
potentials depending on the parity of $n$. For $n$ even, the periodic
superpotential $\beta_1$ does not appear as a separate term in 
(\ref{final}), affecting
only one of denominators. The resulting susy
partners have only a finite quantity of singularities. This fact 
has been used by Stalhofen [8] by constructing potentials 
with bound states embedded in the continuum. On the other hand, 
for $n$ odd, the function $\beta_1$ is a separate term in the 
sum (\ref{final}) and its global effect is not canceled 
by any similar term. The corresponding susy 
partners become singular periodic potentials.

In conclusion, the nonlinear difference algorithm (\ref{solucionk}) allows
the construction of higher order susy partners of any initial potential
$V_0(x)$, provided that a certain number of solutions of (\ref{riccati1})
have been given. This finite difference algorithm generalizes the
superposition principle reported in [5] extending its applications to the
susy construction of new solvable potentials. In particular, the higher
order susy partners $V_n(x,\epsilon_n)$ of the free particle potential
represent a wide set of transparent wells in the terms discussed in
[7--9], as well as multisoliton solutions of the KdV equation as given in
[5].

\vspace{2mm}


This work was performed under the auspices of CONACyT (Mexico) and DGES
project PB94-1115 from Ministerio de Educaci\'on y Cultura (Spain), as
well as by Junta de Castilla y Le\'on (CO2/197). BM and ORO acknowledge
the kind hospitality at Departamento de F\'{\i}sica Te\'orica, Universidad
de Valladolid (Spain). ORO wishes to acknowledge partial finantial support
from the ICSSUR'99 Organizing Committee. 

\vspace{5mm}

{\large \bf References}

{\footnotesize
\begin{itemize}

\item[{[1]}]
D.J.~Fern\'andez~C, {\em Int. J. Mod. Phys. A} {\bf 12}, 171 (1997); \\
D.J.~Fern\'andez~C., M.L.~Glasser and L.M.~Nieto, {\em Phys. Lett. A} {\bf
240}, 15 (1998)

\item[{[2]}]
D.J.~Fern\'andez~C., V.~Hussin and B.~Mielnik, {\em Phys.
Lett. A} {\bf 244}, 1 (1998);\\
J.O.~Rosas-Ortiz, {\em J. Phys. A} {\bf 31}, L507 (1998); {\em J. Phys. A}
{\bf 31}, 10163 (1998);\\
D.J.~Fern\'andez~C. and V.~Hussin, {\em J. Phys. A} {\bf 32}, 3603 (1999) 

\item[{[3]}]
B.~Mielnik, L.M.~Nieto and O.~Rosas-Ortiz, {\em The finite difference
algorithm for higher order supersymmetry}, Universidad de Valladolid {\em
Preprint}, Spain (1998)

\item[{[4]}]
V.~E. Adler, {\em Physica D} {\bf 73}, 335 (1994)

\item[{[5]}]
H.~D. Wahlquist and F.~B. Estabrook, Phys. Rev. Lett. {\bf 31}, 1386
(1973) 

\item[{[6]}]
C.~S. Gardner, J.~M. Greene, M.~D. Kruskal and R. Miura, {\em Phys. Rev.
Lett.} {\bf 19}, 1095 (1967)

\item[{[7]}]
V.B.~Matveev and M.A.~Salle, {\em Darboux Transformations and
Solitons}, Springer-Verlag, Berlin (1991)

\item[{[8]}]
A.~Stahlhofen, {\em Phys. Rev. A} {\bf 51}, 934 (1995)

\item[{[9]}]
B.N.~Zakhariev and V. M. Chabanov, {\em Inverse Problems} {\bf 13}, 
R47 (1997)

\end{itemize}
}

\end{document}